\documentclass[aip,pop,author-year,reprint]{revtex4-1}
\usepackage{amsmath,amssymb}
\renewcommand{\vec}{\mathbf}

\newcommand{\J}{~\mathrm{J}}
\newcommand{\W}{~\mathrm{W}}
\newcommand{\cm}{~\mathrm{cm}}
\newcommand{\kms}{~\mathrm{km/s}}
\newcommand{\G}{~\mathrm{G}}
\newcommand{\nm}{~\mathrm{nm}}
\newcommand{\eV}{~\mathrm{eV}}
\newcommand{\EE}{\mathbf{E}}
\newcommand{\BB}{\mathbf{B}}
\newcommand{\UU}{\mathbf{U}}
\newcommand{\byc}{B_y^{\mathrm{(cos)}}}
\newcommand{\bys}{B_y^{\mathrm{(sin)}}}
\newcommand{\bzc}{B_z^{\mathrm{(cos)}}}
\newcommand{\bzs}{B_z^{\mathrm{(sin)}}}

\newcommand{\ie}{\emph{i.~e.\ }}
\newcommand{\eg}{\emph{e.~g.\ }}
\usepackage{graphicx}
\usepackage{xcolor}

\begin{document}
\title{Hybrid simulations of a parallel collisionless shock in the Large Plasma Device}
\date{\today}
\author{Martin~S.~Weidl}
\affiliation{Department of Physics \& Astronomy, University of California, Los Angeles, California 90095, USA}
\author{Dan~Winske}
\affiliation{Los Alamos National Laboratory, Los Alamos, New Mexico 87545, USA}
\author{Frank~Jenko}
\author{Chris~Niemann}
\affiliation{Department of Physics \& Astronomy, University of California, Los Angeles, California 90095, USA}
\begin{abstract}
We present two-dimensional hybrid kinetic/magnetohydrodynamic simulations of planned laser-ablation experiments in the Large Plasma Device (LAPD). Our results, based on parameters which have been validated in previous experiments, show that a parallel collisionless shock can begin forming within the available space. Carbon-debris ions that stream along the magnetic-field direction with a blow-off speed of four times the Alfv\'en velocity excite strong magnetic fluctuations, eventually transfering part of their kinetic energy to the surrounding hydrogen ions. This acceleration and compression of the background plasma creates a shock front, which satisfies the Rankine--Hugoniot conditions and can therefore propagate on its own. Furthermore, we analyze the upstream turbulence and show that it is dominated by the right-hand resonant instability.
\end{abstract}
\maketitle

\section{Introduction}

Since collisionless shocks are closely connected to the notoriously complicated subject of kinetic plasma turbulence, it is no wonder that much about their physics is still barely understood, such as their formation, the diffusion and acceleration of ions, and the dynamics of electrons at the shock \citep{lembege2004selected}. Satellite measurements of the Earth's bow shock showed already in the 1960s \citep{fairfield1967magnetic} that the width of the actual shock transition between two different magnetohydrodynamic equilibrium states can be a fraction of the collisional mean free path of ions, yet with a turbulent foreshock extending over more than ten earth radii \citep{fairfield1969bowshock}, as had been speculated by theorists \citep{sagdeev1966cooperative}. Even today, the mechanisms that allow magnetic turbulence to grow fast enough to scatter incoming particles and to dissipate energy at a sufficient rate (up to $10~\mathrm{W~m}^{-3}$ \citep{wilson2014quantified}) have not been satisfactorily resolved.

Hybrid simulations, which model the kinetic physics of ions but use an inertialess-fluid model for electrons, have made great progress in explaining satellite observations that could not be understood from the limited data gathered by one-point (or, at best, four-point \citep{lucek2002cluster}) measurements alone. For the quasi-parallel case, in which the mean magnetic field is aligned with the shock-normal direction, a series of non-linear simulations in the 1980s confirmed previous analytical predictions \citep{parker1961quasilinear,golden1973ion} that three ion/ion-beam instabilities were involved in the shock-formation process: the right-hand resonant instability (RHI), the left-hand resonant instability (LHI), and the non-resonant instability (NRI) \citep{gary1984electromagnetic,winske1985coupling,winske1986electro}. As the formation of a quasi-parallel shock takes place over hundreds of ion inertial-lengths, these early simulations had to be performed with low resolution and dimensionality and in a frame of reference co-moving with the shock front. 

One-dimensional simulations \citep{omidi1990steepening} revealed how RHI-induced fast magnetosonic waves can develop into high-amplitude magnetic pulsations upstream of the shock front, which can trap beam ions or background ions \citep{terasawa1988nonlinear,akimoto1991steepening}. As these pulsations are convected into the shock front, they are converted into Alfv\'en waves commonly associated with the NRI \citep{kraussvarban1991structure}, as has also been observed in three-dimensional, low-resolution simulations of the Earth's bow shock \citep{lin2005threedimensional}. Dense fast-ion beams can favor the growth of the NRI over the RHI already upstream of the shock front if the phase-space configuration is sufficiently unstable with respect to the firehose mode \citep{quest1988theory}. Yet the dispersion of the beam ions and the faster propagation of Alfv\'en waves relative to the beam velocity imply that, under solar-wind conditions, fast magnetosonic waves achieve a greater amplitude than Alfv\'en waves \citep{onsager1991interaction}. On the other hand, both one- and two-dimensional simulations \citep{terasawa1988nonlinear,dubouloz1995twodimensional} showed that the heating of ion bunches trapped in pulsations can cause the LHI mode to become unstable and even dominate in parts of the upstream region.

The latter simulations also indicated that the differential deceleration of these pulsations near the shock creates a multidimensional patchwork of magnetic-field structures, as opposed to an easily localizable shock front. This model had previously been suggested by \citet{schwartz1991quasiparallel} on the basis of satellite data which showed that the magnetic field is often bent in a quasi-perpendicular direction close to the Earth's bow shock \citep{schwartz1992observations}. The implication of this framework is, of course, that a full understanding of quasi-parallel shock physics cannot be obtained from simulations with reduced dimensionality.

However, an easily accessible three-dimensional collisionless shock can be created with a high-energy laser in a laboratory, as Dawson already proposed in 1964 \citep{dawson1964production}. After numerous theoretical concept studies \citep{takabe1999recent,remington1999modeling,drake2000design,zakharov2003collisionless}, the field of high-energy-density laboratory astrophysics (HEDLA) has recently produced first experimental results, \eg, the observation of perpendicular collisionless shocks \citep{niemann2014observation} and of the Weibel filamentation \citep{huntington2015observation}.

In this article, we present two-dimensional hybrid simulations and show that the formation of a parallel collisionless shock can be studied in a laboratory experiment like the Large Plasma Device (LAPD) at UCLA \citep{constantin2009collisionless,gekelman2016upgraded}. After explaining the design of the proposed experiment and the numerical methods we employ in Section~\ref{secSetup}, we discuss a series of three runs with increasing beam-ion density and velocity in Section~\ref{secRuns}. The turbulence that develops in the last run is further analyzed in Section~\ref{secTurbulence}, revealing properties remarkably similar to the previously cited bow-shock simulations. Section~\ref{secDiscussion} contains a discussion of the implications and the validity of these results for the planned experiment, before we summarize our conclusions in Section~\ref{secConclusion}.

\section{Experimental and numerical setup}
\label{secSetup}
\subsection{Experimental background}

The two main components of the collisionless-shock experiment are the Large Plasma Device (LAPD), which confines a column of hydrogen plasma that is 0.6 meters in diameter and 17 meters long, permeated by a 200-Gauss magnetic field, and the high-energy Raptor laser system ($E\leq 250\J$ with a wavelength of $1053\nm$).    

When the laser beam is focused to intensities around $10^{12}\W\cm^{-2}$ on a polyethylene (C$_2$H$_4$) target positioned within the LAPD plasma, debris ions (primarily C$^{4+}$) are ablated and fly off into the hydrogen plasma at a normal angle to the target surface at velocities of several $100\kms$. If the ions propagate perpendicularly to the magnetic field, as shown in \citet{clark2013hybrid,niemann2014observation,schaeffer2014laser}, the carbon debris can compress the ambient hydrogen plasma via Larmor coupling sufficiently to form a perpendicular collisionless shock. Directing the laser debris cloud along the background field makes it possible to study the formation of a quasi-parallel shock.

With the help of magnetic-flux probes positioned along the length of LAPD, one can obtain spatially and temporally resolved profiles of the three-dimensional magnetic field. The dynamics and spatial distribution of the debris and ambient ions can be measured spectroscopically and with Langmuir probes.

\subsection{Numerical model}

The simulations which we present in the following are based on the same hybrid model that has been used previously to reproduce and explain measurements of perpendicular shocks at LAPD \citep{clark2013hybrid,clark2014enhanced}. The phase space of hydrogen and carbon ions is discretized using a particle-in-cell (PIC) approach, whereas electrons are modelled as a massless neutralizing MHD fluid ($n_e = n_H + Z_C n_C$, with $Z_C=4$). The electric field is computed at each time step from the momentum equation for the electrons as
\begin{equation}
    \EE = \frac{\left(\nabla \times \BB\right) \times \BB}{e\ n_e} - \frac{\UU_i \times \BB}{c} - \frac{\nabla P_e}{e\ n_e},
    \label{eqnEdef}
\end{equation}
where $\UU_i$ is the charge-weighted flow velocity of both ion species combined and the electron pressure is derived from a polytropic equation of state, $P_e \propto n_e^{5/3}$. Using an alternative model for the electron pressure, \eg, an isothermal- or cold-electron model, would be equally justifiable since no experimental data is available yet to validate either model.

The magnetic field is initialized as a homogeneous field $B_0 = 200\G$ directed along the $x$-axis and evolves according to the Maxwell-Faraday equation, using sub-cycling when necessary to achieve the desired accuracy. On the other hand, the time step which a Boris pusher uses to advance the quasiparticles representing ions is held fixed at $\Delta t = 5\cdot10^{-3}~\omega_g^{-1}$, where $\omega_g$ is the proton gyrofrequency in the magnetic mean field.

All fields are computed and stored on a two-dimensional grid with periodic boundary conditions and physical dimensions of $L_x \times L_y = 2048~\delta_i \times 32~\delta_i$. Convergence tests have shown that {for the purposes of this study} it is sufficient to resolve one inertial length $\delta_i = c/\omega_p=9~\mathrm{cm}$ with four grid cells. {This resolution limits the spectral range over which wave turbulence may develop; so when we refer to turbulence in the following, we mean a relatively narrow spectrum of electromagnetic waves with amplitudes large enough that non-linear interactions may be important.}

We initialize the hydrogen plasma with a uniform density $n_0 = 6\cdot10^{12}~\mathrm{cm}^{-3}$ and a temperature of $T_i = 1.0~\mathrm{eV}$. With these parameters, the length of LAPD (17 meters) corresponds to about 190 inertial lengths {or, for protons traveling at the Alfv\'en speed of $v_A = 178~\mathrm{km~s}^{-1}$, to about 1.5 collisional mean free paths. Since a possible shock propagates super-Alfv\'enically by definition, we can safely speak of a collisionless plasma.} The electron temperature starts out with $T_e = 6.0~\mathrm{eV}$. The initial properties of the carbon-debris ions follow below, case by case.

\section{Discontinuity formation}
\label{secRuns}
\subsection{Sub-Alfv\'enic compression}
\label{subSubalf}

\begin{figure}[htp]
\centering
\includegraphics[scale=.90]{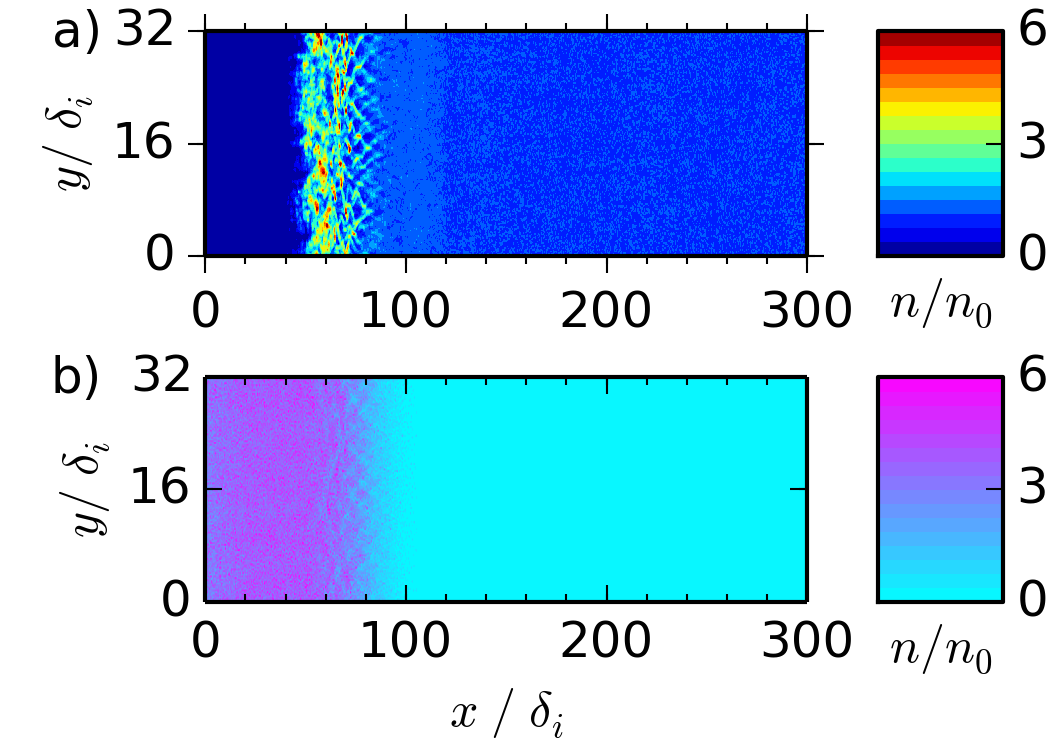}
\caption{Density distribution of a) the hydrogen background plasma, b) the carbon-debris ions for the sub-Alfv\'enic run of subsection \ref{subSubalf} at time $\omega_g t = 50$}
\label{figV1d20}
\end{figure}
\begin{figure}[htp]
\centering
\includegraphics[scale=.90]{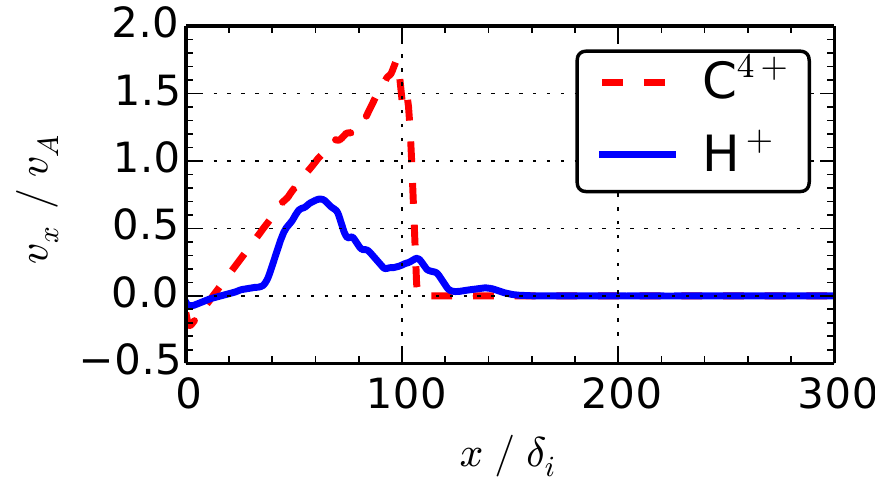}
\caption{Velocity profiles of the hydrogen background plasma (solid) and the carbon debris ions (dashed), averaged over the $y$--direction, at $\omega_g t = 50$}
\label{figV1v20}
\end{figure}

\begin{figure}[htp]
\centering
\includegraphics[scale=.90]{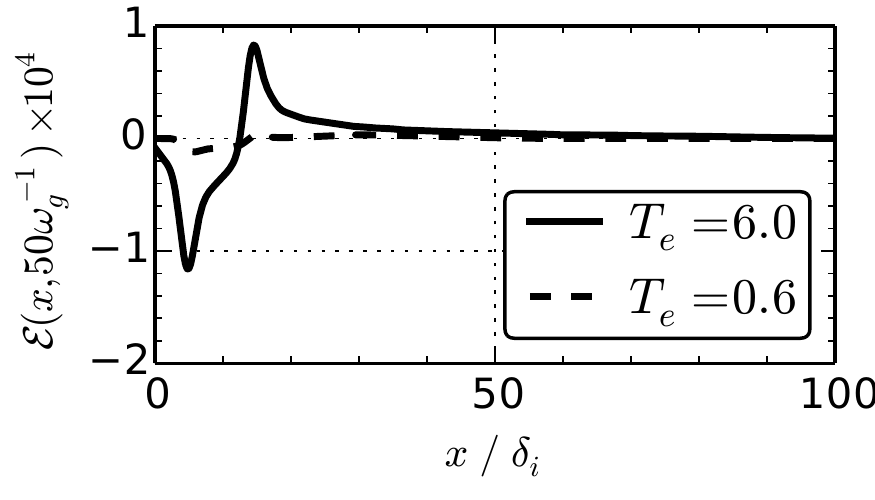}
\caption{Averaged electric field $\mathcal E$ (see definition~(\ref{eqnEMF})) after 50 inverse gyrofrequencies, for runs with electron temperature $T_e = 6.0\eV$ (solid) and $T_e = 0.6\eV$ (dashed)}
\label{figV1e20}
\end{figure}

\begin{figure}[htp]
\centering
\includegraphics[scale=.90]{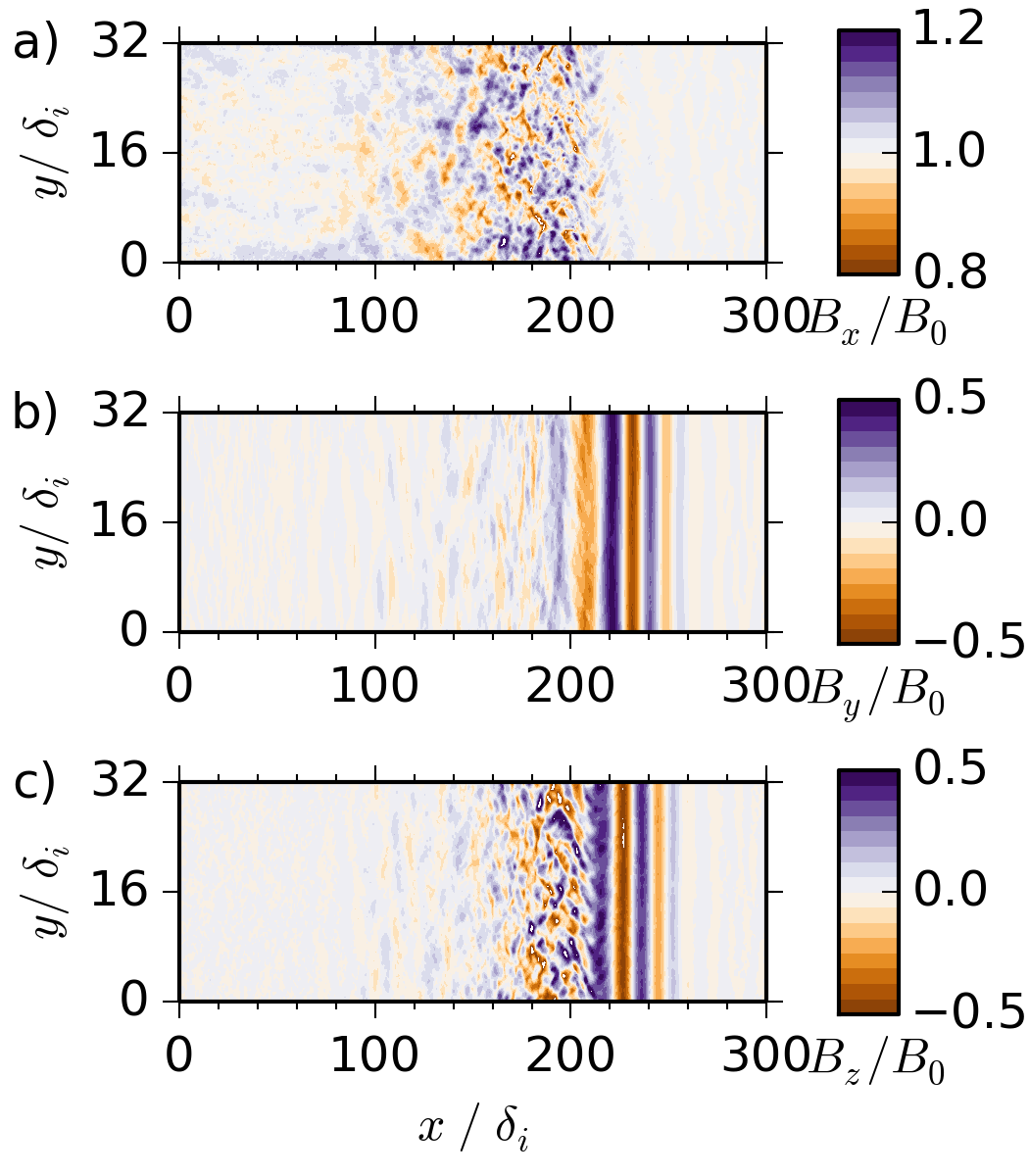}
\caption{Magnetic-field profiles of each field component at $\omega_g t = 130$}
\label{figV1b52}
\end{figure}

In the first simulation we investigate how an initially sub-Alfv\'enic debris population affects the hydrogen plasma. We initialize the debris ions with a density of $n_C = 24~n_0$, uniformly distributed in the interval $5~\delta_i < x < 14~\delta_i$, and a velocity of only $v_0 = 0.5\ v_A$. This mean drift velocity is directed along the positive $x$--axis and thus along the magnetic field direction on average, although a random angular deviation of $\pm15^\circ$ is superimposed to simulate the fact that the blow-off from the target is not entirely unidrectional. As an additional random component, the initial debris-ion velocity is modified with an isotropically distributed thermal component ($T_i = 1\eV$).

Figure~\ref{figV1d20} shows the density distributions of both ion populations after the debris ions have propagated to the right for 50 gyroperiods. The debris is spread over the entire region $x < 80~\delta_i$ and has heated the background plasma sufficiently to form a compression front in the region $50~\delta_i < x < 80~\delta_i$, where the average hydrogen density is about 50~\% higher than the initial value, although locally compression factors of up to 9.0 can be observed. This compression is made possible by acceleration of the hydrogen bulk to $v \approx 0.7~v_A$ (Figure~\ref{figV1v20}).

To identify what causes this acceleration, we compare the parallel electric-field component $E_x$ to the electric field in a run with the initial electron temperature reduced by ten ($T_e = 0.6\eV$). Figure~\ref{figV1e20} contains plots of the quantity
\begin{equation}
\mathcal E(x,T) = \int_0^{L_y} dy \int_0^T dt~E_x(x,y)~\frac{\omega_g}{L_y~B_0},
\label{eqnEMF}
\end{equation}
the normalized parallel electric field averaged over the transverse dimension and time, for both medium- and low-temperature electrons. In the latter run, the electric field is almost ten times weaker than in the fiducial 6-eV~scenario. This strong dependence on the electron temperature confirms that, at least in these early stages of the field evolution, it is the electron-pressure term that dominates the generalized Ohm's law~(\ref{eqnEdef}).

The electron pressure has also accelerated the debris ions closest to the discontinuity to a super-Alfv\'enic velocity and decelerated the left-most ions to negative velocities (Figure~\ref{figV1v20}). Across the whole simulation domain, $B_x$ has developed fluctuations of about 10~\% around its initial value $B_0$ because of the thermal motion of the background plasma. The position of the debris-density discontinuity and the compressed hydrogen at $\omega_g t = 50$ is marked by strong electric currents and hence by oscillations comparable to $B_0$ in all three magnetic-field components.

\begin{figure}[htp]
\centering
\includegraphics[scale=.90]{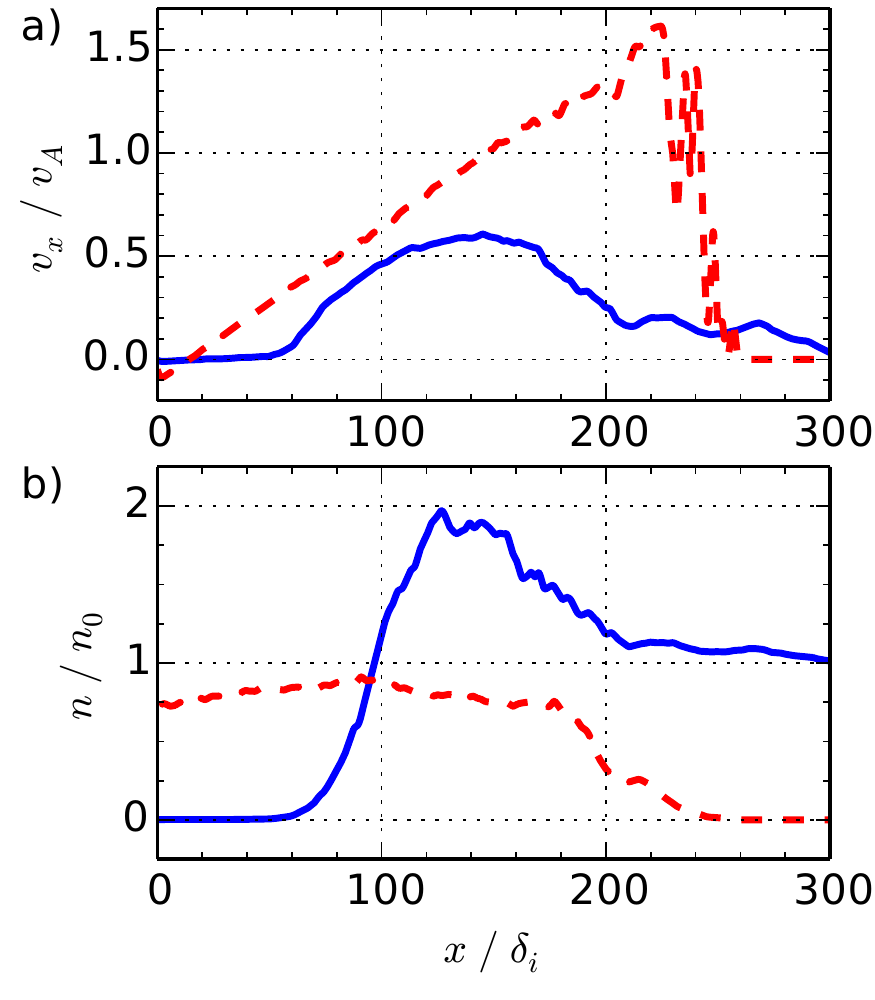}
\caption{Profiles of a) the $x$--velocity, b) the density of both ion populations at $\omega_g t = 130$}
\label{figV1vd52}
\end{figure}

Since the fluctuations of the perpendicular components propagate faster than the fastest carbon ions, they outrun the density discontinuity and the $B_x$--fluctuations despite the electron-pressure acceleration. Thus, 80 inverse gyrofrequencies later (Figure~\ref{figV1b52}), the most pronounced oscillations in $B_y$ and $B_z$ with a wavelength of about 20 inertial lengths appear farther right ($x > 200~\delta_i$) than do the strongest fluctuations of $B_x$ ($100~\delta_i < x < 200~\delta_i$). Where these two regions meet, the spatial structure of the out-of-plane component $B_z$ may be indicative of a developing transverse or oblique mode. The magnetic component $B_x$ is amplified only where the hydrogen density is also elevated, in a region that extends to about 100 inertial lengths behind the front edge of the debris-ion bulk. As a comparison of Figures~\ref{figV1v20} and \ref{figV1vd52}a shows, not much net energy transfer has taken place between the carbon and the hydrogen ions, although the debris ions have decelerated from a top speed of $1.8~v_A$ to $1.6~v_A$. Apart from this reduction and a general time-of-flight broadening, the velocity profiles of both species have not changed much between both figures.

Since the hydrogen plasma is not being accelerated further, its density profile will not steepen any more than depicted in Figure~\ref{figV1vd52}b at later times. It exhibits a gradual increase for about 80 inertial lengths downstream (left) of the debris-density discontinuity to $n \approx 1.9~n_0$, before it drops off relatively quickly (over about 50 inertial lengths) to 0.3~\% of the initial background density. In both the velocity and the density profile of the background plasma, this moderate hump is far too smooth to be considered a shock.

\subsection{Mach-2 compression}
\label{subMach2}

To investigate how an increase in density and velocity affects these results, we have performed a second run with an initial Alfv\'enic Mach number of $M_A = 2.0$. In addition to quadrupling the velocity of the carbon ions, we have raised their initial density by the same factor to $n_C = 96~n_0$. Keeping the total number of debris ions approximately equal, we have reduced the width of the region initially populated with carbon to a quarter its previous value. All other parameters, like the initial temperature, maintain their values as described above. This configuration is essentially identical to the one used by \citet{clark2013hybrid} to reproduce experimental measurements of perpendicular collisionless shocks at LAPD.

\begin{figure}[htp]
\centering
\includegraphics[scale=.90]{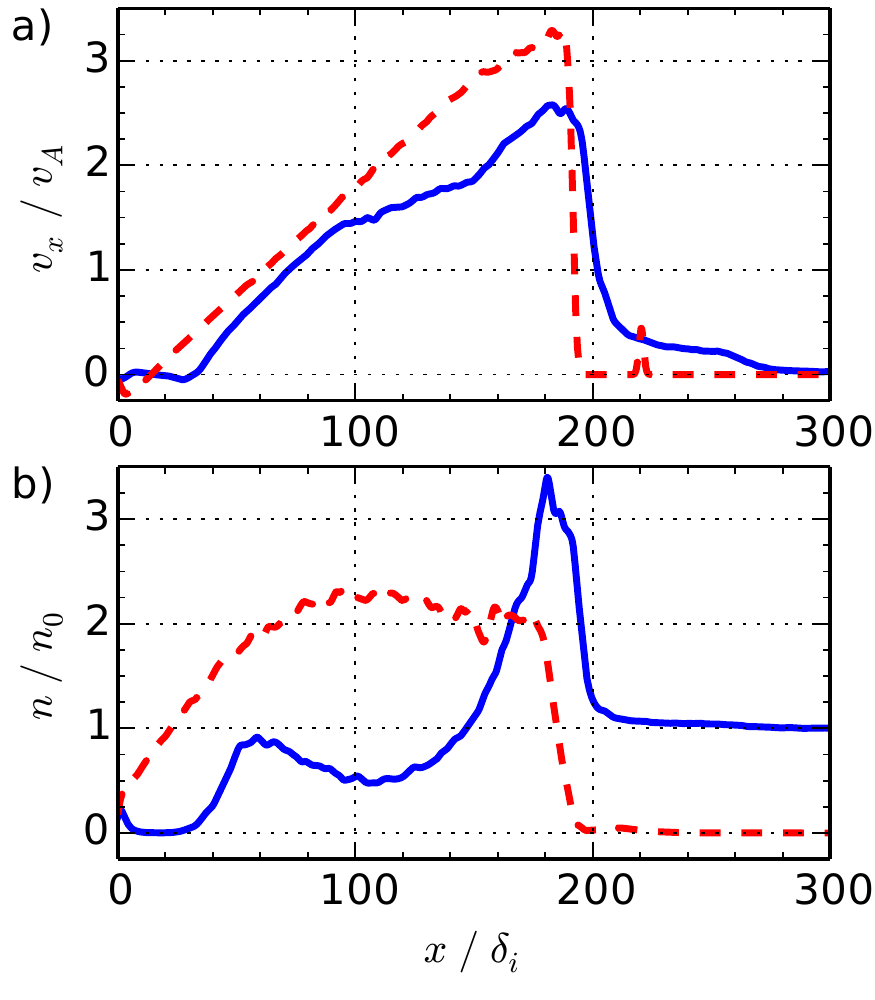}
\caption{Profiles of a) the $x$--velocity, b) the density of the hydrogen ions (solid) and the carbon ions (dashed) for the $M_A = 2$ run of subsection \ref{subMach2} at $\omega_g t = 50$}
\label{figV2vd20}
\end{figure}

After 50 inverse gyrofrequencies, the position of the front edge of the debris cloud coincides with a sharp peak in the velocity profiles of both species and in the hydrogen-density profile (Figure~\ref{figV2vd20}). Whereas the velocity profiles of carbon and hydrogen have assumed similar sawtooth-like shapes, the density profiles exhibit important differences with respect to both each other and the sub-Alfv\'enic case of Figure~\ref{figV1vd52}b: Immediately upstream of a broad plateau in the debris-density distribution ($50~\delta_i < x < 180~\delta_i$), the hydrogen ions are compressed into a density spike peaking at $n \approx 3.5~n_0$, or about 150~\% of the debris-density plateau behind it, with a steep increase over only 20 inertial lengths. For comparison, in the previous subsection the hydrogen density had begun ramping up slowly just downstream of the front edge of the debris cloud and had taken 80 inertial lengths to reach its peak. The initially fourfold density of the debris cloud, which after thermal expansion results in a plateau that is still twice as high as in the sub-Alfv\'enic scenario, accelerates the background ions much more efficiently and thus pushes them ahead of itself like a piston. As indicated by the similar velocity profiles of both species (and shown below in Figure~\ref{figV2bperp}), this configuration will propagate stably for several hundred inertial lengths without breaking up.

\begin{figure}[htp]
\centering
\includegraphics[scale=.90]{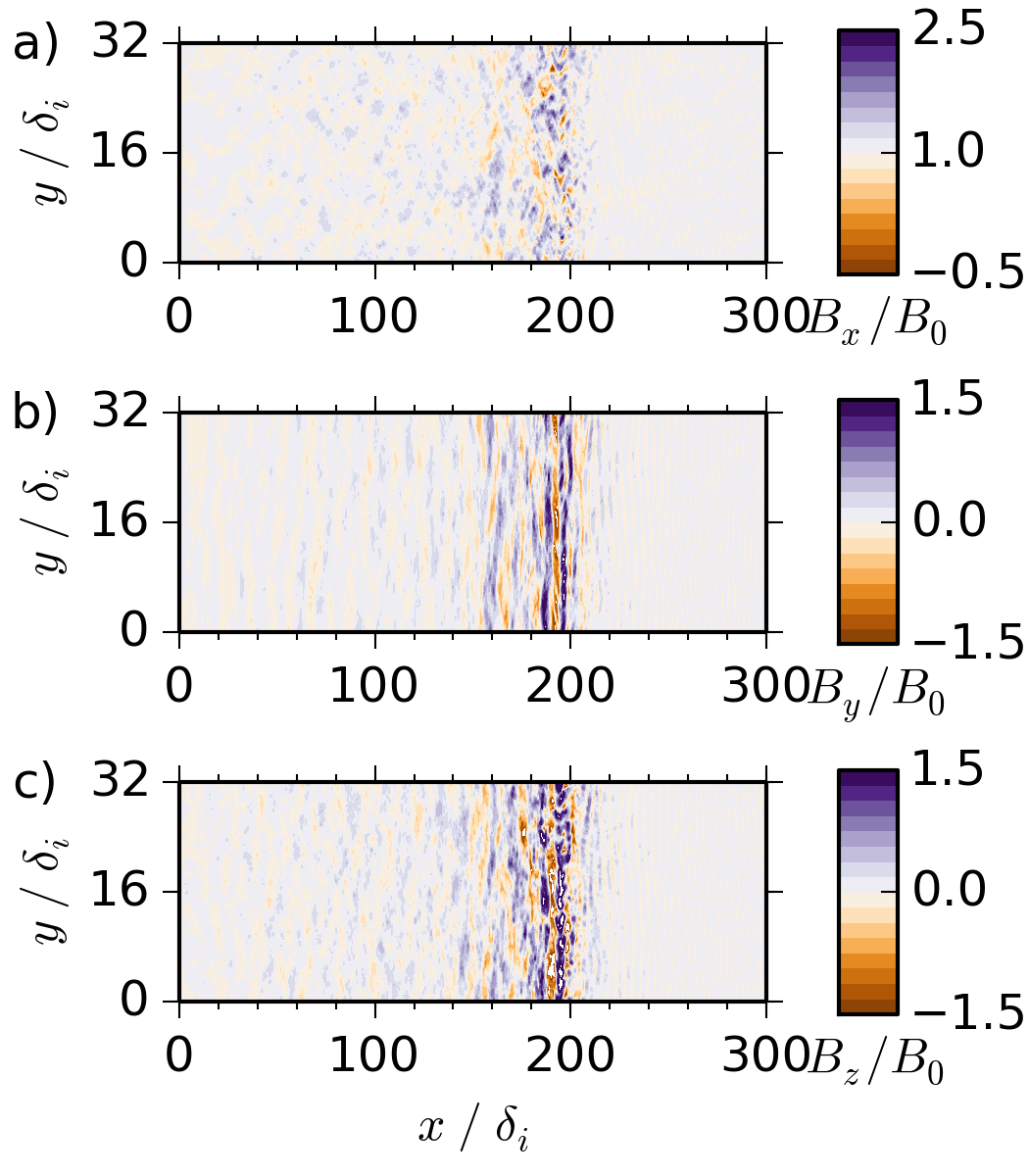}
\caption{Components of the magnetic field in the a) $x$--, b) $y$--, c) $z$--direction at time $\omega_g t = 50$}
\label{figV2b20}
\end{figure}

The magnetic field at the same time is depicted in Figure~\ref{figV2b20}. The large velocity of the carbon ions prevents the dominant fluctuations in the perpendicular components from overtaking the parallel fluctuations. Only the low-amplitude, short-wavelength oscillations which are barely visible in the right of the $B_y$-plot have a sufficient phase velocity. Hence, the magnetic field upstream of the density discontinuity ($x > 200~\delta_i$) is essentially unperturbed, in contrast to the sub-Alfv\'enic case. Moreover, the amplitude of the dominant oscillations is up to three times larger.

\begin{figure}[htp]
\centering
\includegraphics[scale=.86]{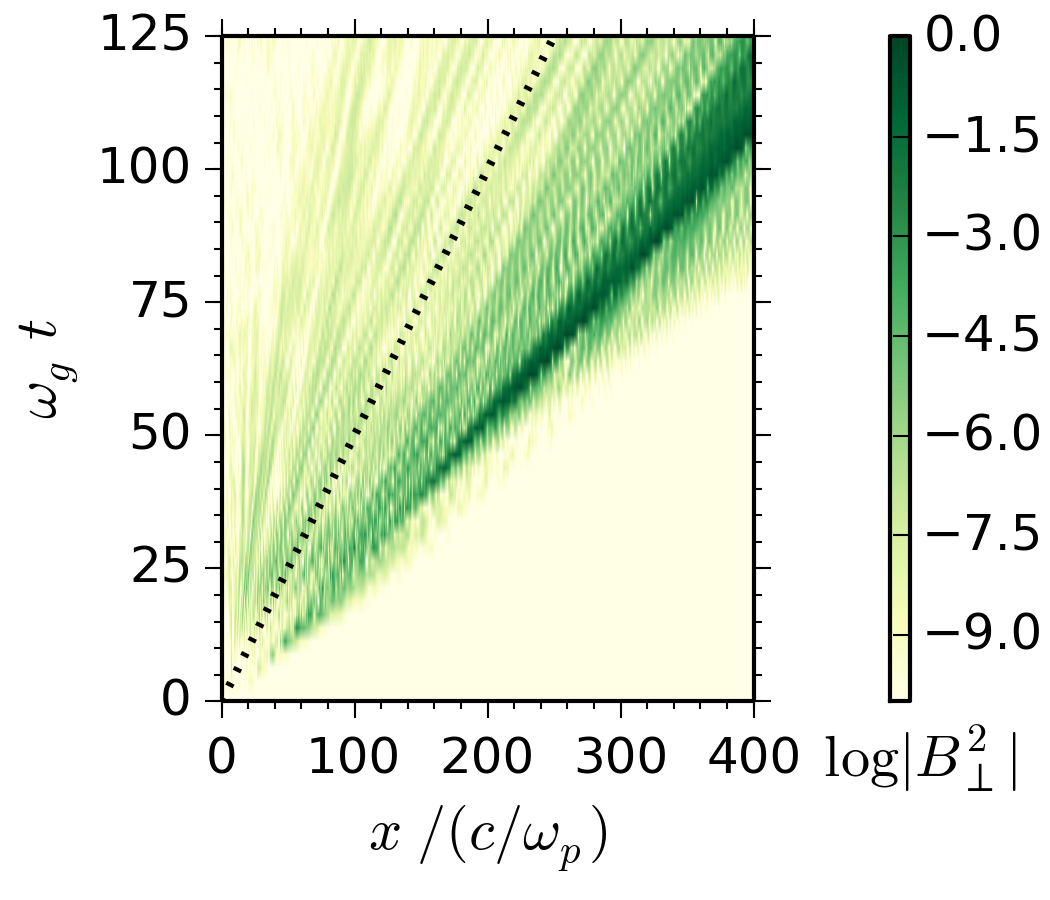}
\caption{Evolution of the perpendicular magnetic-field strength (arb.~u.) in the $M_A = 2$ run of subsection \ref{subMach2} (the dotted line indicates $v=2~v_A$)}
\label{figV2bperp}
\end{figure}

An overview of the perpendicular magnetic-field intensity as it evolves over 125 inverse gyrofrequencies is presented in Figure~\ref{figV2bperp}. It is clear that the magnetic perturbations propagate significantly faster than not only the Alfv\'en velocity, but also the initial velocity $v_0 = 2~v_A$ of the debris ions (indicated by the dotted line). It is therefore not the bulk of the debris cloud, which is centered around $x = 100~\delta_i$ in Figure~\ref{figV2vd20}b, but the few most energetic carbon ions, quickly reaching $v = 4~v_A$ due to thermal expansion and electron-pressure heating, that heat, accelerate, and thus compress the background hydrogen. This efficient energy transfer, in turn, decelerates the fastest debris ions to about three times the Alfv\'en velocity and leads to the sharp debris-density discontinuity just behind the hydrogen-density peak. Both compression fronts, for hydrogen as for carbon, are much steeper than the more subdued increases of the corresponding densities towards their plateaus in the sub-Alfv\'enic run shown in Figure~\ref{figV1vd52}b.

Returning to Figure~\ref{figV2bperp}, one easily sees that the magnetic turbulence caused by the hydrogen compression trails off quickly downstream of the debris-density discontinuity and has all but disappeared where the bulk of the carbon is located. Thus, the magnetic compression cannot be due to the presence of the carbon ions themselves, but only due to the sharply localized density increase of the hydrogen. However, this compression is not sufficiently self-sustaining to be called a shock. To remedy this problem, a mechanism to excite turbulence ahead of the compression front is needed.

\subsection{Mach-4 shock formation}
\label{subMach4}

As both the density and the initial velocity of the debris cloud were increased in the previous subsection, it is not clear which of the effects described above are due to which parameter. In order to settle this question, we have performed a third run, taking the lower initial density of the carbon debris from the sub-Alfv\'enic run ($n_C = 24~n_0$) and the smaller initial spatial extent from the Mach-2 case. To compensate for the smaller number of carbon ions, we have doubled their initial Alfv\'enic Mach number to 4 so that the total kinetic energy of the debris piston stays the same, as do all remaining parameters.

\begin{figure}[htp]
\centering
\includegraphics[scale=.90]{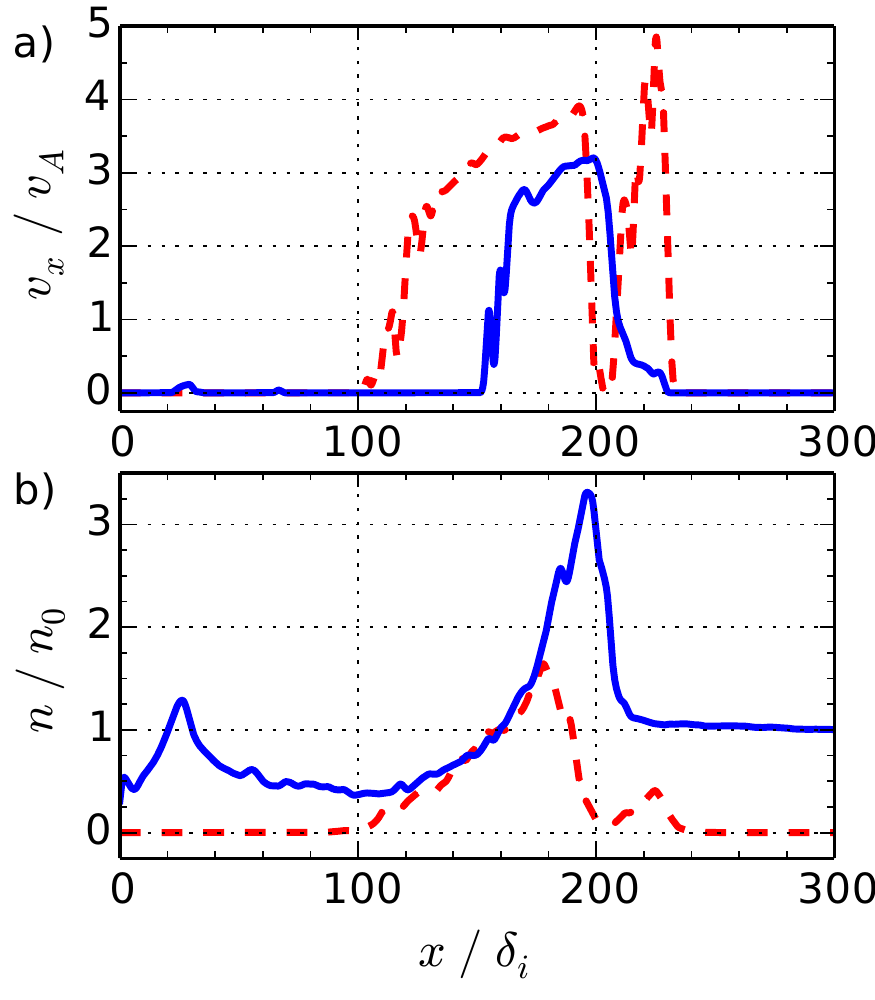}
\caption{Profiles of a) the $x$--velocity, b) the density of the hydrogen ions (solid) and the carbon ions (dashed) for the Mach-4 run of subsection \ref{subMach4} at $\omega_g t = 42.5$}
\label{figV4vd17}
\end{figure}

By the time most of the carbon ions reach the end of the 190 inertial lengths that correspond to the length of LAPD, their velocity and density distribution has developed a bimodal shape (Figure~\ref{figV4vd17}): While their bulk is still peaked around $x = 180~\delta_i$ with a maximum velocity of $v \approx 4.0~v_A$, a smaller second bunch of carbon ions has already advanced to about $x \approx 220~\delta_i$. Early on, the electron pressure accelerated this thin slice of debris ions to up to $v \approx 5.5~v_A$ at $\omega_g t = 30$ so that it could detach from and propagate ahead of the slower bulk, but in Figure~\ref{figV4vd17} this `vanguard' has decelerated to just under $5.0~v_A$.

\begin{figure}[htp]
\centering
\includegraphics[scale=.90]{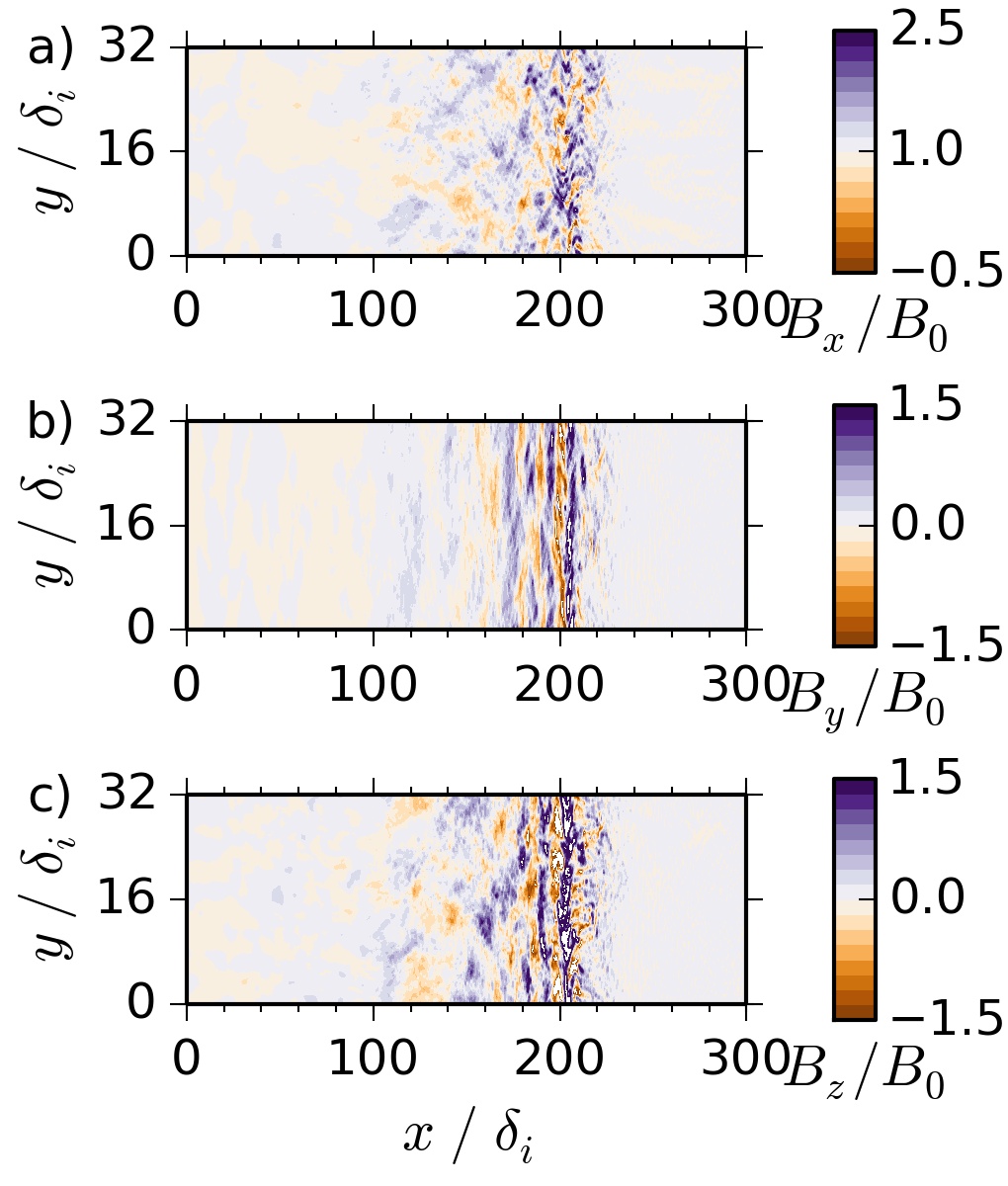}
\caption{Components of the magnetic field in the a) $x$--, b) $y$--, c) $z$--direction at time $\omega_g t = 42.5$}
\label{figV4b17}
\end{figure}

Part of this energy has been transferred to the background plasma, by accelerating the hydrogen ions to $v \approx 3.0~v_A$ and compressing them into a narrow density peak in between the two debris-ion bunches. Another significant fraction of the fast-carbon kinetic energy has gone into perturbing the magnetic field, as shown in Figure~\ref{figV4b17}. Unlike in the Mach-2 case (Figure~\ref{figV2b20}), the magnetic turbulence is amplified even upstream of the hydrogen peak ($200~\delta_i < x < 220~\delta_i$) by the `vanguard' of fast debris ions. 

\begin{figure*}[htp]
\centering
\includegraphics[scale=.90]{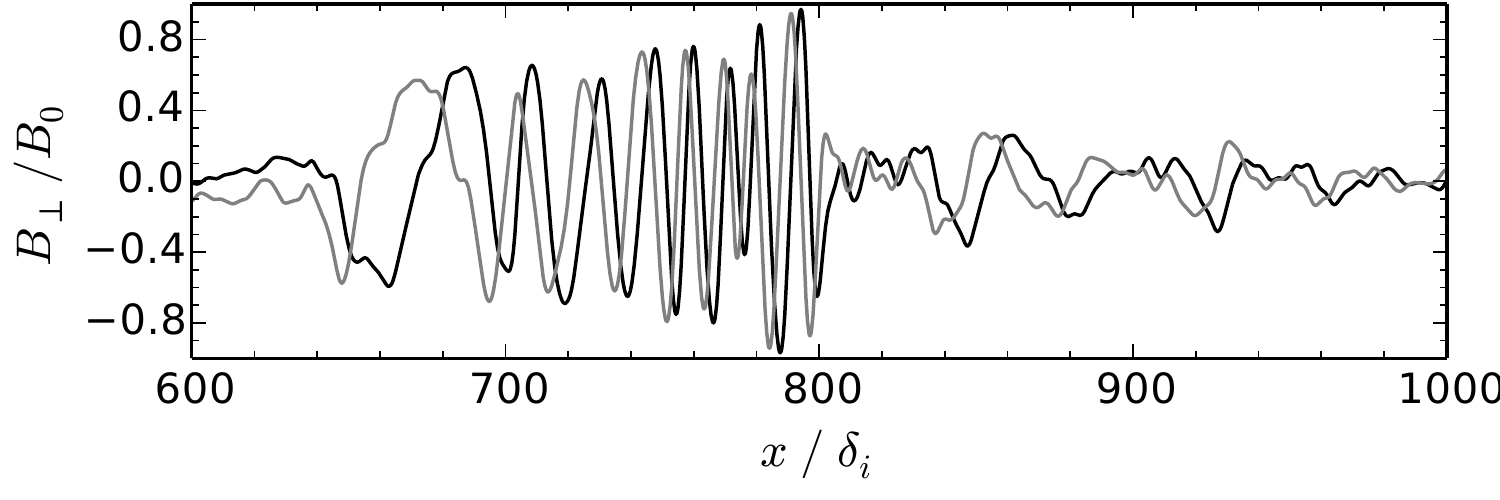}
\caption{Profiles of the magnetic-field components in the $y$-- (black) and $z$--direction (gray) at time $\omega_g t = 200$}
\label{figV4byz80}
\end{figure*}

Together with the debris ions, the area over which this magnetic turbulence is visible quickly spreads from the propagating discontinuity towards both the upstream and the downstream direction, although the turbulence amplitude rises rapidly behind the discontinuity. The result can be seen in Figure~\ref{figV4byz80}, which depicts the components of the magnetic field after 200 inverse gyrofrequencies: Whereas the upstream turbulence ($800~\delta_i < x$) is dominated by a mode with a wavelength of $\lambda \approx 40~\delta_i$ (a more detailed analysis of which follows in the next section) and peaks at $B_y \sim B_z \approx 0.25~B_0$, in the downstream medium ($x < 800~\delta_i$) the dominant wavelength contracts to $\lambda \approx 10~\delta_i$ and the perpendicular field strength quadruples, attaining local values of $B_y \approx 1.0~B_0$. As we prove below, the discontinuity has now unarguably evolved into a shock front.

\begin{figure}[htp]
\centering
\includegraphics[scale=.86]{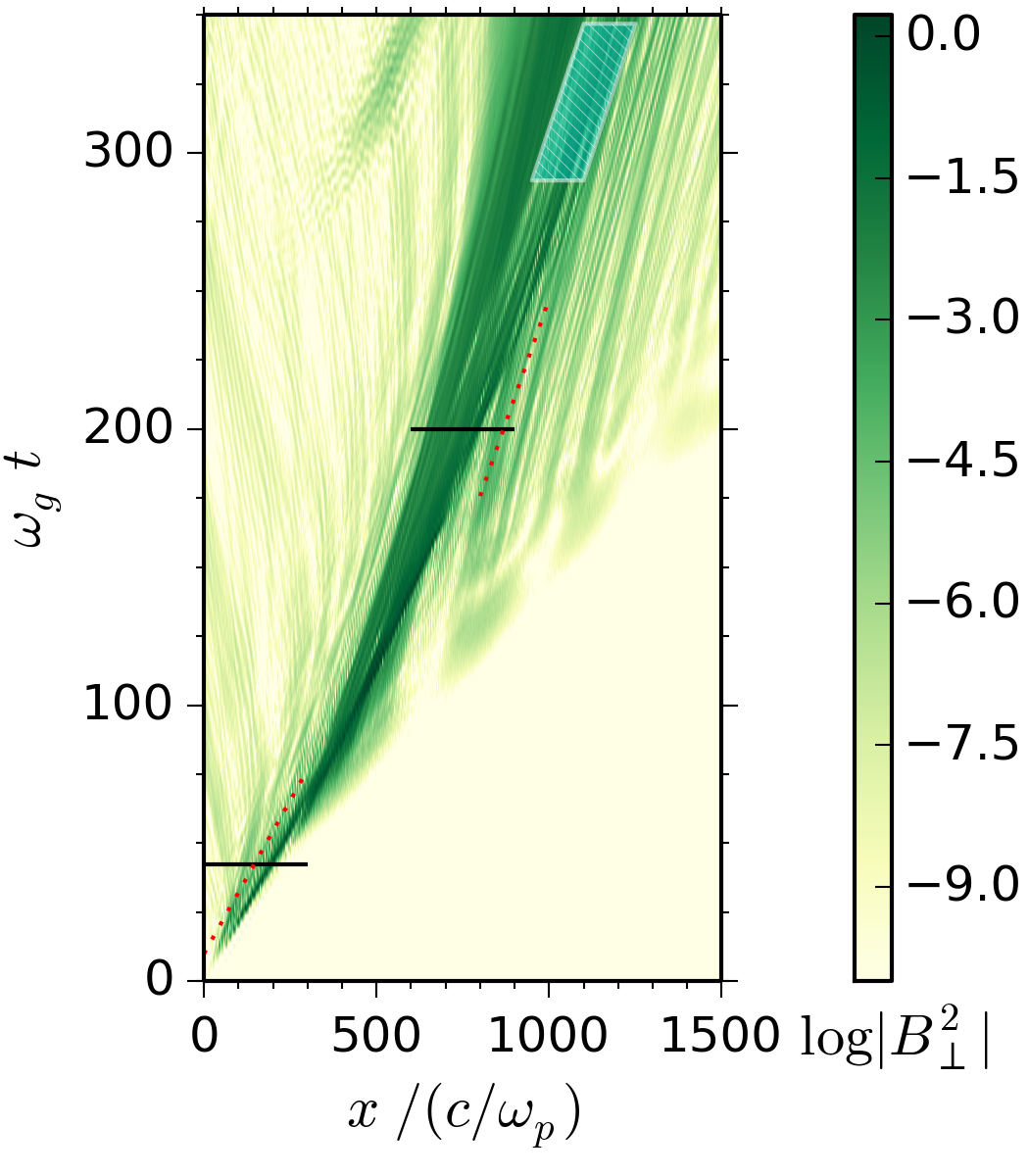}
\caption{Evolution of the perpendicular magnetic-field strength (arb.~u.) in the $M_A = 4$ run of subsection \ref{subMach4} (the cyan parellelogram indicates the region analyzed in section~\ref{secTurbulence})}
\label{figV4bperp}
\end{figure}

Before we move on to discuss the density profiles, Figure~\ref{figV4bperp} shows the evolution of the perpendicular magnetic-field strength. While the fastest perturbations propagate through the plasma with the velocity of the fastest vanguard ions, $v \approx 7~v_A$, they decay into smaller wave packets with group velocity $v_g \approx 2~v_A$. The position of the largest magnetic compression, however, changes at a velocity that is close to $v \approx 5~v_A$ for the first 300 inertial lengths, faster than the bulk of either ion species (Figure~\ref{figV4vd17}a), like in the Mach-2 case, but comparable to the average vanguard-ion speed. After about 500 inertial lengths, as more and more hydrogen ions are swept up, this velocity begins to decrease noticeably while still remaining highly super-Alfv\'enic. After 1000 inertial lengths, it has slowed down to $2.6~v_A$.

Even though the development of the shock presented here occurs beyond the length of LAPD, we emphasize that the main characteristics of a collisionless shock, a sharp discontinuity between two equilibrium configurations of the magnetohydrodynamic quantities, are already present in the system in Figure~\ref{figV4vd17} and are therefore likely to be achieved at LAPD. Indeed, if we take $v_s = 4.5~v_A$ for the shock velocity at $\omega_g t = 42.5$, as indicated by the dotted line in the lower left corner of Figure~\ref{figV4bperp}, the peak values in Figure~\ref{figV4vd17} fulfill the Rankine--Hugoniot conditions for a parallel MHD shock exactly: $n_u (v_{x,u} - v_s) = n_d (v_{x,d} - v_s)$, where $u$ and $d$ denote upstream and downstream equilibrium values for the hydrogen plasma, respectively. An analysis of the energy flux proves that it is also continuous across the surface $x = 200~\delta_i$. Although these relations only describe conservation laws, they confirm that both sides of the shock front are in hydrodynamic equilibrium{, at least during the time in which one can safely define a constant shock-propagation speed}. The downstream region this early in the shock-formation process is only about twenty inertial lengths wide, but given enough time, this structure will grow into a full-blown shock.

\begin{figure}[htp]
\centering
\includegraphics[scale=.90]{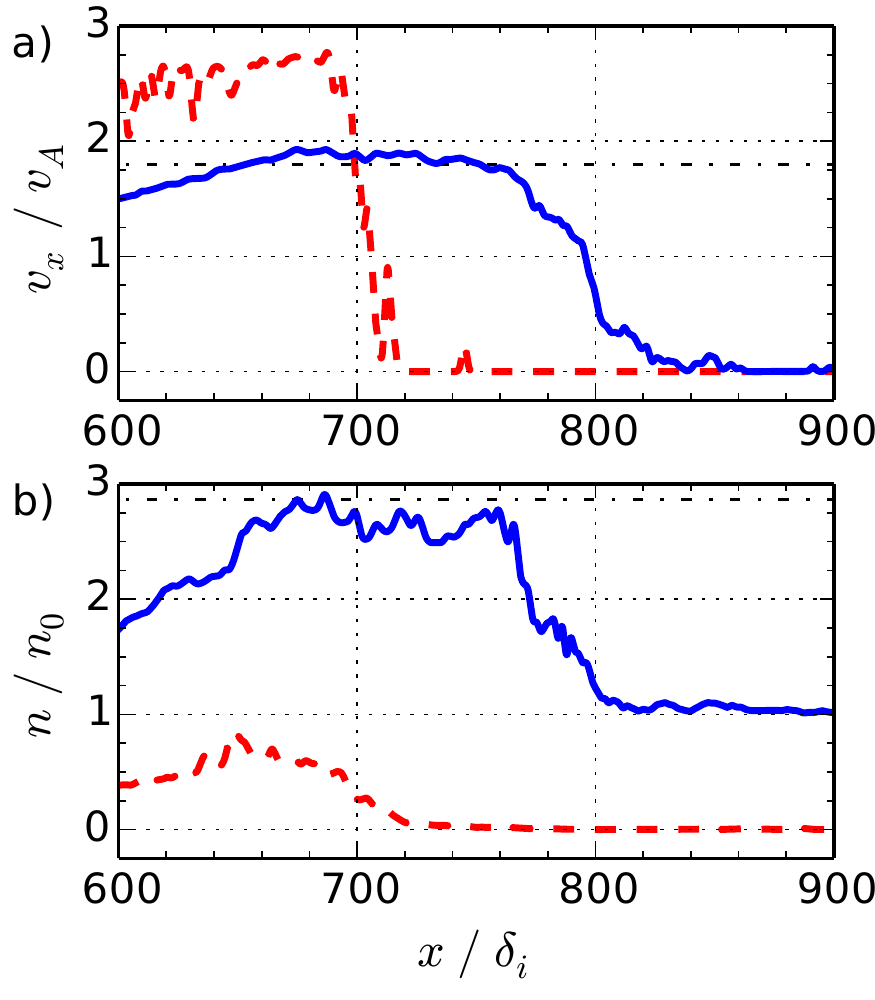}
\caption{Profiles of a) the $x$--velocity, b) the density of the hydrogen ions (solid) and the carbon ions (dashed) at $\omega_g t = 200$, and dash-dotted lines indicating downstream values compatible with $M_A = 2.8$ and the Rankine--Hugoniot conditions for a parallel shock}
\label{figV4vd80}
\end{figure}

This fact is evident from Figure~\ref{figV4vd80}, which contains the velocity and density profiles of both ion species at $\omega_g t = 200$. The hydrogen profiles exhibit a rapid jump at $x = 800~\delta_i$ in correspondence with the magnetic field depicted in Figure~\ref{figV4byz80}. The downstream region is now over a hundred inertial lengths wide, which may be interpreted as the shock having detached from the one remaining debris-ion bunch and now propagating independently of the carbon piston. Moreover, assuming a shock velocity of $v_s = 2.8~v_A$, we find that the MHD variables on both sides of the shock front are in excellent agreement with the Rankine--Hugoniot conditions for a parallel (or hydrodynamic) shock.

Hence, depending on how wide a downstream region one desires before calling the discontinuity a shock, the conditions at LAPD allow the experimental observation of either a parallel collisionless shock or the early stages of its formation. For the geometry we have studied, it appears advantageous to use the available laser energy to accelerate fewer debris ions to higher energies: Although the velocity and density profiles in Figures~\ref{figV2vd20} and \ref{figV4vd17} may look rather similar, the slower propagation speed of the discontinuity in the Mach-2 case means that the hydrogen distribution has not reached a hydrodynamic equilibrium configuration, as can be checked easily with the Rankine--Hugoniot conditions. Instead of more and more hydrogen being compressed to the same density and thus forming a proper shocked medium, the hydrogen-density peak will broaden and eventually disappear. 

The denser Mach-2 bunch also stays more or less coherent, unlike the Mach-4 bunch, which separates into the slower piston bunch and the vanguard bunch of fast debris ions. On one hand, these high-speed ions increase the electron pressure upstream of the hydrogen-density peak and thus prevent it from broadening towards the upstream too early; on the other hand, they also create sufficient upstream turbulence to pre-accelerate the hydrogen before the slower bulk of the carbon arrives.

\section{Shock-precursor turbulence}
\label{secTurbulence}

If we are to call the discontinuity predicted for the Mach-4 scenario a collisionless shock, a strict requirement is that the compression and acceleration of the hydrogen ions is actually caused by interactions with the upstream magnetic turbulence. Thus, a closer analysis of the wave spectrum in the transition region is in order.

\subsection{Theory of left- and right-hand instability}

The turbulence driven by the vanguard debris-ions in the upstream region could be the result of either of two possible ion/ion beam instabilities \citep{gary1985electro}: the left-hand resonant instability (LHI) or the right-hand resonant instability (RHI). The free energy for both originates in the resonant wave-particle coupling between MHD waves and the carbon debris. A fast magnetosonic wave in the background hydrogen plasma that propagates in the same direction as the debris can interact with it, be amplified by resonant ions, and develop into the RHI. Alternatively, if the thermal velocity spread of the carbon-ion population is large enough, individual debris ions that stream opposite to the bulk velocity of the carbon cloud (\ie, to the left) can lose energy to backward-propagating shear-Alfv\'en waves and thus drive the LHI.

Already at $\omega_g t = 200$, the phase relation of $B_y$ and $B_z$ in Figure~\ref{figV4byz80} indicates that the upstream perturbations of the perpendicular field are of predominantly positive helicity. This observation on its own is compatible with both a right-traveling RHI mode and a left-traveling LHI mode \citep{gary1991electro}, however, and therefore cannot replace a detailed Fourier analysis.

Both modes are described \citep{gary1985electro} by the dispersion relation
\begin{equation}
 \omega^2 - k^2 c^2 + \sum_{s\in\{\mathrm H,\mathrm C\}} \omega_{p,s}^2\ \zeta_s^0\ Z(\zeta_s^{\pm1}) = 0,
 \label{eqnDR}
\end{equation}
where $\omega_{p,s}$ denotes the plasma frequency of either ion species, over which the sum runs, and $Z$ is the Fried--Conte plasma dispersion function with the argument
\begin{equation}
 \zeta_s^m = \frac{\omega - k_x v_{x,s} + m\ \omega_{g,s}}{\sqrt2\ k_x\ v_{\mathrm{th},s}},
\end{equation}
with $\omega_{g,s}$ the gyrofrequency and $v_{\mathrm{th},s}$ the thermal velocity of either ion population. The RHI is described by the above relation with $m = +1$, whereas the LHI corresponds to $m = -1$.

For large debris velocities ($v_0 \gg v_A$) and low temperatures ($v_{\mathrm{th},C} \ll v_0$), the configuration may also become unstable to the non-resonant ion/ion beam instability \citep{winske1986electro}. The fluctuations caused by this instability are simple Alfv\'en waves, like those caused by the LHI. To model the influence of the fastest debris ions at the front edge in the following, we use a beam velocity of $v_b = 2.6~v_A$.

\begin{figure}[ht]
\centering
\includegraphics[scale=.90]{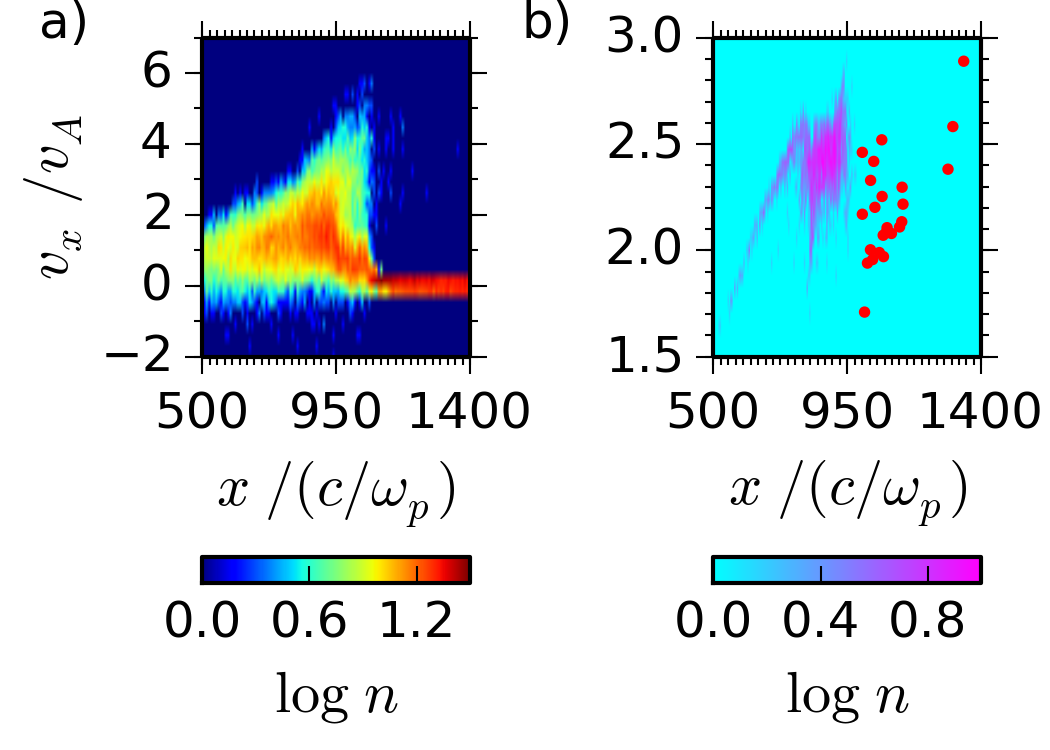}
\caption{Densities of the $x$--$v_x$--phase space (arb.~u.) of the a) hydrogen ions and b) carbon ions in the Mach-4 run at $\omega_g t = 290$. The red circles denote the phase-space location of sample vanguard debris-ions}
\label{figV4psd29}
\end{figure}

\subsection{Moving-window dispersion analysis}

The patch of turbulence that we have studied in detail is marked with a cyan parallelogram in Figure~\ref{figV4bperp}. Choosing this particular interval is mainly motivated by our goal of analyzing a region that contains strong magnetic fluctuations with almost constant propagation speed, which facilitates a Fourier analysis with unambiguous results, yet does not contain more carbon ions than are needed to drive the beam instabilities. The mechanism we want to elucidate is how a few vanguard debris-ions seed sufficient turbulence that the bulk of the hydrogen ions upstream of the shock is pre-heated and pre-accelerated to form a shock precursor (Figure~\ref{figV4psd29}a). The red circles in Figure~\ref{figV4psd29}b indicate the phase-space location of individual quasi-particles that represent these vanguard ions; the phase-space density of the vanguard bunch is too low to be visible in the pseudocolor plot without these extra markers. Note that the upstream magnetic turbulence has scattered the vanguard ions close to the shock front perpendicularly and thus reduced their parallel velocity.

\begin{figure}[htp]
\centering
\includegraphics[scale=.86]{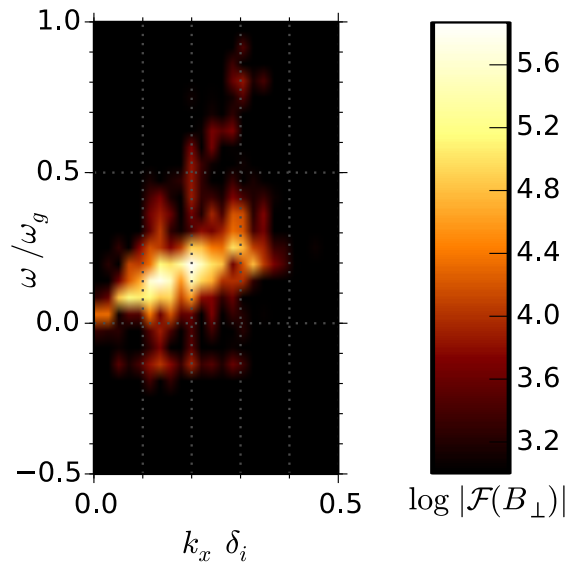}
\caption{Power spectrum of the upstream turbulence in a moving window of width $\Delta x = 150~c/\omega_p$, as measured in the moving-window frame}
\label{figPower}
\end{figure}

\begin{figure}[htp]
\centering
\includegraphics[scale=.86]{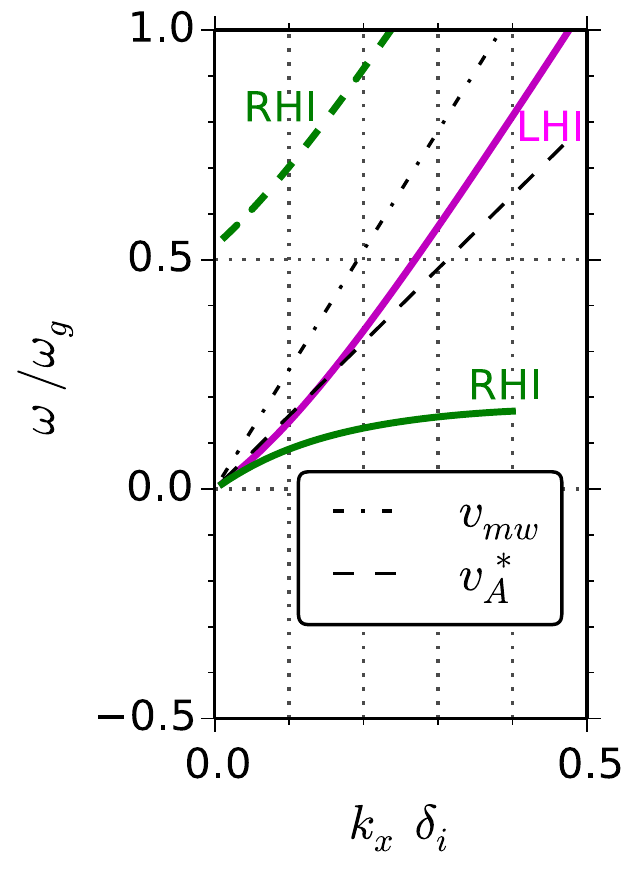}
\caption{Dispersion relations of the right-hand and left-hand instability for a beam density $n_b = 0.3~n_0$, in the frame moving with the shock-front velocity $v_{mw} = 2.6~v_A$. Also shown: moving-window velocity $v_{mw}$, which would correspond to the apparent velocity of an object fixed in the lab frame, and the Doppler-shifted Alfv\'en velocity $v_A^\ast$}
\label{figDispB}
\end{figure}

To determine the dominant wave mode, we first transform the data in the selected interval into a frame of reference that moves with the shock velocity $v_{mw} = 2.6~v_A$. When viewed in this reference frame, the magnetic turbulence is quasi-stationary, in the sense that the amplitude of the oscillations in $B_y$ and $B_z$ at a fixed point in space is almost constant. In other words, the dominant modes group-stand in this frame. Only then is the data Fourier-transformed along the $x$-- and $t$--dimensions to determine the wavelengths and frequencies of the dominant modes. Figure~\ref{figPower} shows the resulting power spectrum of the perpendicular magnetic field in the $k_x\,$--$\,\omega\,$--plane. For comparison, Figure~\ref{figDispB} provides the graphs of the analytic dispersion relation~(\ref{eqnDR}) with the beam parameters extracted from the phase-space data ($v_\mathrm{th}/v_A = 1.0$ and $0.3$ for hydrogen and carbon, respectively). In the latter plot, the Doppler shifting of the frequency axis into the moving-window frame has already been accounted for.


A comparison of Figures~\ref{figPower} and \ref{figDispB} reveals that the dominating instability is the RHI. The source of the weaker fluctuations extending diagonally upwards above the RHI branch is ambiguous; the LHI, the NRI, and the moving-window transformation are possible causes. Since all vanguard ions maintain a large rightward velocity and thus cannot couple to a left-traveling Alfv\'en wave, the LHI is by far the least likely candidate. A numerical artifact caused by changing into the rest frame of the shock is a more plausible explanation.

It is also possible that these fluctuations signify Alfv\'en waves excited by the NRI, most likely close to the shock transition. A similar analysis for the downstream medium, defined as the 200 inertial lengths behind the marked interval, confirms previous predictions \citep{kraussvarban1991structure} that the RHI waves are shifted to the Alfv\'en-wave branch as they cross downstream of the shock, indicating that the higher density in the compressed medium favors the NRI.

Further confirmation of the importance of the RHI in the upstream medium can be obtained by decomposing magnetic field profiles parallel to the $x$-axis (at $y_s = 16~\delta_i$) into components with positive and negative helicity, as explained in Appendix~\ref{secHelicity}. Figure~\ref{figBhel} shows the real parts of $B^+(x)$ and $B^-(x)$, the $y$--projections of the positive- and negative-helicity components of the magnetic field, respectively, during the period in which the moving-window analysis was performed.

The positive-helicity component (Fig.~\ref{figBhel}a) contains a large-amplitude wave packet traveling to the right. This indicates that the magnetic field is mainly right-hand circularly polarized during this period, probably because the RHI is being driven by the vanguard ions. The wave speed of fluctuations decreases with longer wavelength, as illustrated by the long-wavelength envelope shifting from the center of the shaded area at $\omega_g~t = 290$ to its left edge at $\omega_g~t = 350$. Meanwhile, the shorter-wavelength oscillations close to the right edge are only slightly slower than the shock front. The negative-helicity component, shown in Fig.~\ref{figBhel}b, exhibits some left-hand polarized fluctuations close to the shock front, further evidence that the NRI provides a larger amount of non-resonant coupling between background and beam ions where the density of the latter increases.

\begin{figure}[htp]
\centering
\includegraphics[scale=0.9]{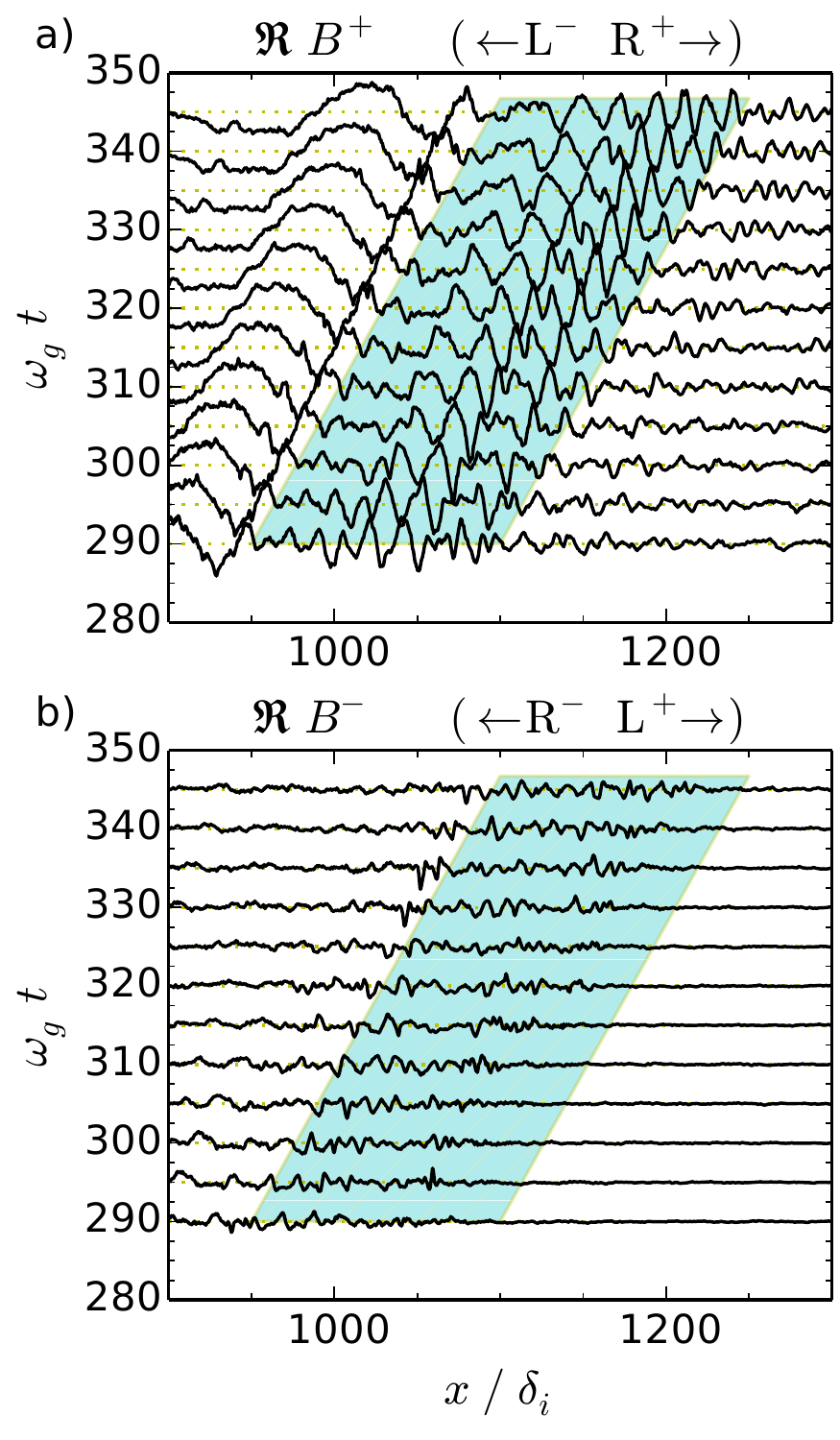}
\caption{Real parts of the components of the magnetic field with \textit{a)}~positive and \textit{b)}~negative helicity with respect to the $x$-axis}
\label{figBhel}
\end{figure}

\section{Discussion}
\label{secDiscussion}

Similar to the Earth's bow shock \citep{barnes1970theory}, the upstream turbulence and the shock-precursor formation at LAPD will be dominated by the RHI. Individual high-energy debris-ions streaming ahead of the bulk of the carbon piston will couple to fast magnetosonic waves in the upstream hydrogen-plasma and amplify the magnetic turbulence. Given enough amplitude, this turbulence will pre-accelerate the background ions to super-Alfv\'enic velocities and compress them accordingly, even before the debris piston advances far enough to have an immediate effect on the hydrogen. From these results, we expect future experiments at LAPD to provide valuable insights into the formation of parallel collisionless shocks.

Although we have shown that the energies and densities that can be achieved at LAPD allow a collisionless shock to form, the limited length of 17 meters may restrict observations to the early stages of the formation process. In this regime, the magnetic turbulence compressing the hydrogen plasma is not yet propagating independently of the carbon-debris piston; instead, it is still confined to the space between the bulk and the vanguard ions of the carbon debris (Figure~\ref{figV4vd80}). It is questionable whether or not detachment from the piston in a parallel-shock configuration at moderate Mach numbers is even possible, however. Previous studies, which used bow-shock parameters in hybrid simulations \citep{omidi1990low}, were able to observe such detachment only in quasi-parallel geometries with a non-zero angle between the shock normal and the magnetic field.

For the sake of semantics, we should point out that the structure we are discussing is categorized as a planar blast wave in the shock physics community \citep{drake2006}. In this jargon, an actual ideal shock would require the piston to maintain a constant pressure so that density and velocity profiles that are, in the shock rest-frame, entirely independent of time can form, with no rarefaction occurring downstream of the discontinuity. This condition is almost impossible to satisfy with any setup based on target ablation, of course.

{%
The transition between the upstream and the downstream region remains well-defined in our simulation once the shock has formed. There is no cyclic disruption and reformation of the shock, as opposed to the one-dimensional simulations of \citet{burgess1989cyclic}, likely because the flow of upstream-generated short-wavelength fluctuations into the shock is relatively steady and not split into individual bursts (see Figure~\ref{figV4bperp}). Burgess attributed the more intermittent arrival of these waves at the quasi-parallel shock in his simulations, which led to its cyclic reformation, to an artificially high electron resistivity. In our simulations, we do not use any anomalous resistivity, which mainly smooths out field gradients transverse to the background magnetic field and would have little bearing on the transverse electromagnetic waves that dominate in the upstream region. However, Burgess also noted that lower-dimensional simulations may overestimate fluctuation amplitudes because three-dimensional space offers more modes into which wave energy can decay. Whether this caveat also applies to our investigation will be determined by comparison with measurements.
}

There are other aspects in which the initial conditions of the actual experiment will be less ideal than what we have assumed in this article. In particular, the initial magnetic field is likely to be perturbed by waves produced by the laser-target interaction or by ions hitting the inner walls of LAPD, which may or may not be conducive to the shock-formation process {(\eg, \citet{giacalone1992hybrid})}. The ablated carbon-ion population will be less of a homogeneous front with a uniform drift velocity, and will instead consist of various charge states with transverse-velocity distributions and a finite transverse size depending on the dimensions of the laser's focal spot. This variation is likely to affect the growth rates of the involved beam instabilities significantly. We are currently researching each of these effects and will discuss their relevance in future publications.

An ion getting reflected multiple times by the turbulent fields on either side of the shock will experience first-order Fermi acceleration and reach a significantly suprathermal energy. Direct observation of Fermi-accelerated particles will be challenging, but measuring the diffusion rate of ions and their dynamics downstream of the shock will allow us to check and possibly improve existing models of diffusive shock acceleration. {To determine whether any particles can possibly be injected into a Fermi-acceleration process at all, we will start by looking for signatures of shock drift acceleration as seen in the simulations by \citet{giacalone1992hybrid}.}

We will also be in a position to verify predictions for the non-linear growth rates of the various instabilities. By identifying the wave modes on both sides of the shock front, quantifying their energy content with magnetic-field probes, and comparing the results with further simulations, we will determine how valid the assumptions which current models make about the formation of parallel shocks actually are.

\section{Conclusion}
\label{secConclusion}

Assuming parameters similar to those successfully used to predict the outcome of perpendicular-shock experiments at LAPD, we have shown that ablated carbon ions with $M_A=4$ can initiate the formation of a collisionless parallel shock. Lower debris velocities may result in a temporary compression of the hydrogen background, provided that the kinetic energy density is still high enough, but this is only a transient phenomenon as opposed to a steady shock.

A low-energy setup that starts out with carbon ions arranged in a planar front that is 9 inertial lengths thick, moving at half the Alfv\'en velocity and with a density of $n = 24~n_0$, ultimately has insufficient energy density to accelerate the hydrogen plasma to more than its original, sub-Alfv\'enic velocity. With all forward-directed waves escaping the compression region into the upstream plasma, the density profile does not steepen very much before the initial  carbon energy is completely dissipated.

Quadrupling both the initial density and the initial velocity of the ablated carbon, we find that the background plasma is significantly compressed and accelerated to $M_A = 3$. However, the compression region is confined to the vicinity of the front edge of the carbon bulk. This piston comprises almost all carbon ions, such that insufficient magnetic turbulence is excited in the upstream plasma.

By doubling the initial debris-ion velocity once more while keeping the kinetic energy constant, a parallel collisionless shock can be produced. The higher velocity allows part of the carbon to stream ahead as vanguard ions, exciting sufficient turbulence to accelerate some of the background plasma, but exerting enough pressure in the early stages of the forming shock to prevent the compressed hydrogen from expanding into the upstream direction and flattening. We have shown that the downstream plateau behind this sharp discontinuity, which satisfies the Rankine--Hugoniot conditions, could extend to over a hundred inertial lengths if it were given enough space, proving that this structure is a self-consistent propagating shock which will be eventually maintained by the upstream turbulence. The majority of the magnetic energy in this regime resides in fast magnetosonic waves driven via the RHI by the beam of vanguard ions. 

Comparing these results to measurements in an upcoming experimental campaign at LAPD will enable us to identify possible shortcomings of the current framework of collisionless-shock formation. Future research will address the individual roles of and the interaction between the RHI and the NRI, the effect of electron dynamics which we have largely neglected so far, and the possibility of using an experiment like LAPD to observe Fermi acceleration.

\begin{acknowledgments}
This work was facilitated by the Max-Planck/Princeton Center for Plasma Physics and supported by DTRA under Contract~No.~HDTRA1-12-1-0024, by DOE under Contract~Nos.~DE-SC0006583:0003 and DE-NA0001995, and by NSF~Award~No.~1414591.
\end{acknowledgments}

\appendix
\section{Helicity decomposition}
\label{secHelicity}

We define a helical fieldline that winds around the $x$-axis to possess positive helicity with wavenumber $k^+ > 0$ if it follows a space curve $\vec s_+(x; k^+, r, \phi_0)$ with
\begin{multline}
\vec s_+(x; k^+, r, \phi_0) =\\ x~\hat{\vec x} + r \left(\cos(k^+ x + \phi_0)~\hat{\vec y} - \sin(k^+ x + \phi_0)~\hat{\vec z}\right),
\end{multline}

for $x \in \mathbb{R}$, $r>0$, and $0 \leq \phi_0 < 2\pi$. Analogously, a helical fieldline which possesses negative helicity with wavenumber $k^- > 0$ can be described by the space curve
\begin{multline}
\vec s_-(x; k^-, r, \phi_0) =\\ x~\hat{\vec x} + r \left(\cos(k^- x + \phi_0)~\hat{\vec y} + \sin(k^- x + \phi_0)~\hat{\vec z}\right).
\end{multline}

For an arbitrary space curve $\vec B(x)$ with $B_x > 0$ everywhere, we define its positive- and negative-helicity components by their Fourier transforms,
\begin{align}
\tilde B^+(k) &= \frac12 \left[ \tilde B_y(k) + i \tilde B_z(k) \right]\label{eqnBpluscmp}\\
              &= \frac{\byc(k) - \bzs(k)}{2} + i \frac{\bzc(k) + \bys(k)}{2},\\
\tilde B^-(k) &= \frac12 \left[ \tilde B_y(k) - i \tilde B_z(k) \right]\label{eqnBminuscmp}\\
              &= \frac{\byc(k) + \bzs(k)}{2} - i \frac{\bzc(k) - \bys(k)}{2}.
\end{align}

for strictly positive $k$, and $\tilde B^+(k) = 0 $ and $\tilde B^-(k) = 0$ for $k \leq 0$. These definitions are equivalent to $B^l$ and ${B^r}^\ast$, respectively, as previously defined by \citet{terasawa1986decay}. Here the Fourier transform, the sine transform, and the cosine transform of $B(x)$ are defined as
\begin{align}
\tilde B(k) &= \int \frac{dx}{\sqrt{2\pi}}~B(x)~e^{i k x},\\
B^{\mathrm{(sin)}}(k) &= \int \frac{dx}{\sqrt{2\pi}}~B(x)~\sin(k x),\\
B^{\mathrm{(cos)}}(k) &= \int \frac{dx}{\sqrt{2\pi}}~B(x)~\cos(k x).
\end{align}

It is easily seen that, for $\vec B(x) \equiv \vec s_+(x; k^+, r, \phi)$, we get
\begin{align}
\tilde B^+(k) &=\sqrt{\frac\pi2}~r~e^{-i \phi}~\delta(k^+ - k),\\
\tilde B^-(k) &\equiv 0.
\end{align}

Likewise, for a space curve $\vec s_-(x; k^-, r, \phi)$ with negative helicity we find
\begin{align}
\tilde B^+(k) &\equiv 0,\\
\tilde B^-(k) &= \sqrt{\frac\pi2}~r~e^{-i \phi}~\delta(k^- - k).
\end{align}

For a fieldline that follows a superposition of multiple space curves with positive or negative helicities, the phase relations between these different modes are conserved in the complex phase of $\tilde B^+$ and $\tilde B^-$. Hence, the spatial structure of the perpendicular components $B_y$ and $B_z$ is conserved by the helicity components and can be approximately reconstructed by an inverse Fourier transform:
\begin{equation}
B^\pm(x) = \int \frac{dx}{\sqrt{2\pi}}~\tilde B^\pm(k)~e^{-i k x},
\end{equation}

with large values of $|\mathfrak{R}\,B^\pm(x)|$ for those $x$-intervals in which $|B_y(x)|$ is large. Since $\tilde B^\pm(k) = 0$ for $k<0$ by definition, $\tilde B^\pm$ are generally not Hermitian functions and $B^\pm$ are complex-valued functions; regions where $|\mathfrak{I}\,B^\pm(x)|$ is large indicate regions where $|B_z(x)|$ is large.

In general, therefore, $x$-intervals in which a fieldline winds around the $x$-axis with a predominantly left-handed sense of rotation, which corresponds to positive helicity in our definition, are marked by large-amplitude oscillations in the graphs of $\mathfrak{R}\,B^+$ and $\mathfrak{I}\,B^+$. If these oscillations propagate in the direction opposite to the $x$-axis as time progresses, the fieldline exhibits predominantly left-handed circular polarization in these intervals. On the other hand, if the oscillations propagate parallel to the $x$-axis, as in Figure~\ref{figBhel}, the fieldline is mainly right-handed circularly polarized.

To compute the helicity decomposition of data that are discretized on a grid of $N$ uniformly spaced points separated by $\Delta$, we perform a fast Fourier transform on the signal and apply equations~(\ref{eqnBpluscmp}) and (\ref{eqnBminuscmp}) on all components with a frequency $0 < k < \frac{\pi}{\Delta}$, with the remaining components set to zero. An inverse FFT then yields the complex positive- and negative-helicity components.

%

\end{document}